\documentclass[12pt]{article}
\usepackage{latexsym, epsfig, graphics}
\newcommand{\be}{\begin{equation}}
\newcommand{\ee}{\end{equation}}
\newcommand{\bq}{\begin{eqnarray}}
\newcommand{\eq}{\end{eqnarray}}
\newcommand{\intst}{\int_{0}^{p^{+}}\! d\sigma \int d\tau}
\newcommand{\intdbl}{\int_{0}^{p^{+}} d\sigma \int d\tau \int_{0}
^{p^{+}} d\sigma' \int d\tau'}
\newcommand{\tfi}{(\tau_{f}-\tau_{i})}
\begin{document}
\begin{titlepage}
\today          \hfill 
\begin{center}
\hfill    LBNL-59617\\

\vskip .5in

{\large \bf Summing Planar Bosonic Open Strings}
\footnote{This work was supported 
 by the Director, Office of Science,
 Office of High Energy and Nuclear Physics, 
 of the U.S. Department of Energy under Contract 
DE-AC02-05CH11231}
\vskip .50in

%alternate footnote for faculty:
%\footnote{This work was supported in part by the Director, Office of 
%Energy Research, Office of High Energy and Nuclear Physics, Division of 
%High Energy Physics of the U.S. Department of Energy under Contract 
%DE-AC03-76SF00098 and in part by the National Science Foundation under 
%grant PHY-0098840.}

\vskip .5in
Korkut Bardakci\footnote{e-mail:kbardakci@lbl.gov}

{\em Department of Physics\\
University of California at Berkeley\\
   and\\
 Theoretical Physics Group\\
    Lawrence Berkeley National Laboratory\\
      University of California\\
    Berkeley, California 94720}
\end{center}
\vskip .5in

\begin{abstract}

In earlier work, planar graphs of massless $\phi^{3}$ theory were summed
with the help of the light cone world sheet picture and the mean field
approximation. In the present article, the same methods are applied to
the problem of summing planar bosonic open strings. We find that 
in the ground state of the system, string boundaries form a condensate
on the world sheet, and a new string emerges from this summation. Its
slope is always greater than the initial slope, and it remains non-zero even
when the initial slope is set equal to zero. If we assume the  initial
string tends to a field theory in the zero slope limit, this result
provides evidence for string formation in field theory.  

\end{abstract}
\end{titlepage}%THIS PAGE (PAGE ii) CONTAINS THE LBL DISCLAIMER

\newpage
\renewcommand{\thepage}{\arabic{page}}
\setcounter{page}{1}
%THIS IS PAGE 1 (INSERT TEXT OF REPORT HERE)
\noindent{\bf 1. Introduction}

\vskip 9pt

A couple of years ago,
the present author and Charles Thorn initiated a program of studying
field theory in the planar limit
 by reformulating it as a local theory on the world sheet [1].
This new formulation provided a fresh approach for tackling some of the
old standing problems. The field theory most intensively studied so far
is massless $\phi^{3}$ theory [2-6], although Thorn and collaborators
later extended the world sheet approach to more realistic models [7-9].
Apart from providing a new insight into field theory, the world sheet
formulation enables one to do dynamical calculations, using the
mean field approximation. The most interesting result so far to come 
out of the mean field method was the emergence of  a string picture
 from the sum of the planar graphs in the $\phi^{3}$ theory. Of
course, there are various caveats: The model is unphysical and in fact
unstable, and  the reliability of the mean field approximation is open
to question. There are also various technical problems which were only
partially overcome in [6]. In spite of all the drawbacks, we feel that
an important step forward has been taken.

In this article, instead of summing field theory diagrams, we consider
the sum of planar bosonic open string diagrams, and we show that the
mean field method is also applicable to this case. We list below the main
 motivations for this generalization:\\
1) Some of the technical problems encountered in summing field
theory diagrams are absent in the case of the string diagrams.
In fact, summing strings turns out to be simpler than summing
$\phi^{3}$ graphs.\\
2) Assuming that the zero slope limit of the string theory
is some field theory, one can indirectly recover the field theory
sum from the string sum by taking the zero slope limit.\\
3) The string sum is of interest in itself; it may enable one to
investigate problems such as tachyon condensation [10].

The main results to emerge from this investigation are the following:
After the summation, a new string emerges, whose slope is greater
than the original slope. The dynamical mechanism for this change is
what we call the condensation of the string boundaries.
What happens is that the boundaries become dense on the world sheet,
changing its texture. This phenomenon was already observed in the
context of the $\phi^{3}$ theory [2-6]; in fact,
in this respect, the string and
field theory calculations are remarkably similar. The crucial point
is that even after the initial slope is set equal to zero, the final slope
after the summation remains finite. Since the zero slope limit of string
theory is generally believed to be a field theory, this result supports
the idea of string formation in field theory.

All of this, of course, depends crucially on the validity of the mean
field approximation. In addition, there are some questions on the 
meaning of the zero slope limit. Usually, in taking the zero slope
limit, the vector meson is kept at zero mass, and the heavy particles
decouple. The resulting field theory is therefore a vector (gauge)
theory. The existence of the tachyon, however, throws some doubt on
this picture; in this limit, the tachyon becomes infinitely heavy
and therefore infinitely destabilizing. Clearly, it is desirable to
generalize the approach developed in this paper to the tachyon free
 superstring theory, where this problem is absent. The present
article can be thought of as a warmup exercise for this future project.

In section 2, we briefly review the Feynman graphs in the mixed lightcone
variables [11] and the and the local field theory on the world sheet
which generates these graphs [1-3]. We also discuss the transformation
properties of various fields under a special Lorentz boost, which
manifests itself as a scale transformation on the world sheet.

 The technology introduced in  section 2 for summing over field theory graphs
will turn out to be exactly what is needed later on for summing over planar
strings in section 3. As it turns out, the action on the world sheet
that reproduces the string sum can be cast in a form very similar
to the corresponding action for field theory by means of a duality
transformation. This action is then the starting point of the mean
field method developed in section 4 from the point of view of the
large $D$ limit, where $D$ is the dimension of the transverse space.
 Part of this section is in the
nature of a review, since there is a lot in common here with
references [2-6], where the mean field method was applied to the
$\phi^{3}$ field theory. The section ends with a discussion of how
to define cutoff independent parameters from the cutoff dependent ones.

 In section 5, the mean field method is applied to the calculation
of the ground state of the model. The equations determining the ground
state have two possible solutions, which we call the $(+)$ and $(-)$ 
phases. The $(+)$ phase describes  the original perturbative sum
of the strings. The $(-)$ phase, which has lower energy and therefore is
the true ground state, is the phase where the string boundaries have formed
a condensate on the world sheet. We show that,
 as a result of this condensation,
a new string is formed, with a slope greater than the slope of the
original string. This new slope remains non zero even when the initial
slope is set equal to zero. Identifying the zero slope limit of the
string with field theory, we consider this result as a strong indication
of string formation in field theory.

Finally, we discuss our results in section 6, and summarize our
conclusions and point out some future directions of research in
section 7.

\vskip 9pt

\noindent{\bf 2. A Brief Review}

\vskip 9pt

 In this section, we present a brief review of the results obtained in
references [1-6]. In this work, starting with the world sheet
representation of the sum of the planar graphs of the
 massless $\phi^{3}$ field theory [1-3], the ground
state energy of the system was calculated in the mean field
approximation. It was found that, subject to this approximation, the 
dynamics favors string formation.

 The starting point of the mean field calculation
 is the light cone representation of the scalar
propagator [11]
\be
\Delta(p)=\frac{\theta(\tau)}{2 p^{+}}\exp\left(-i\tau\, \frac{{\bf p}
^{2}+m^{2}}{2 p^{+}}\right),
\ee
where $p^{+}=(p^{0}+p^{1})/\sqrt{2}$ and $\tau= x^{+}=(x^{0}+x^{1})
/\sqrt{2}$. Here the superscripts $0$ and $1$ label the timelike
and the longitudinal directions, and the transverse momentum ${\bf p}$
lives in the remaining $D$ dimensions. The propagator is represented by
a horizontal strip of width $p^{+}$ and length $\tau$ on the world
sheet (Fig.1).
\begin{figure}[t]
\centerline{\epsfig{file=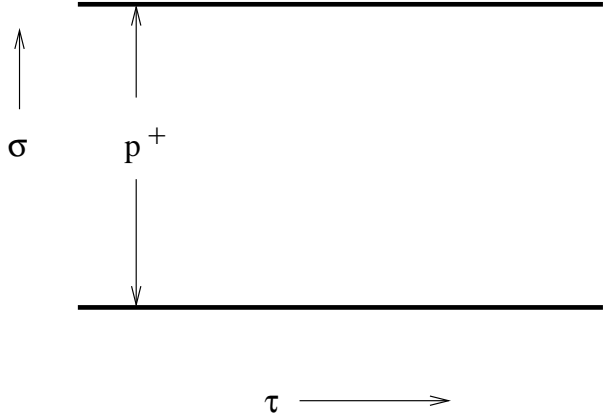,width=8cm}}
\caption{The propagator}
\end{figure}
 The solid lines that form its boundary carry transverse
momenta ${\bf q}_{1}$ and ${\bf q}_{2}$ flowing in opposite directions
so that
$$
{\bf p}= {\bf q}_{1} - {\bf q}_{2}.
$$

More complicated graphs consist of several horizontal line segments
(Fig.2).
\begin{figure}[t]
\centerline{\epsfig{file=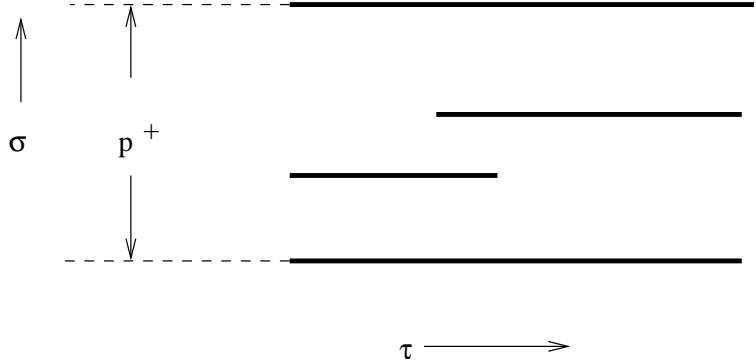,width=10cm}}
\caption{A typical graph}
\end{figure}
 The $\phi^{3}$ interaction takes place at the beginning and at
the end of each segment, where a factor of $g$ (the coupling constant)
is inserted. One  has then to integrate over the position of the
vertices and over the momenta carried by the solid lines.

It was shown in [2,3] that the light cone Feynman rules sketched above
can be reproduced by a local world sheet field theory. The world sheet
is parametrized by the coordinate $\sigma$ along the $p^{+}$
direction and $\tau$ along the $x^{+}$ direction, and the transverse
momentum ${\bf q}$ is promoted to a bosonic field ${\bf q}(\sigma,\tau)$
on the world sheet. In addition, two fermionic fields $b$ and $c$, each with
the $D/2$ components (assuming that $D$ is even), are needed. The action
on the world sheet for the massless theory is given by
\be
S_{0}=\intst \left(b'\cdot c' -\frac{1}{2} {\bf q}'^{2}\right),
\ee
where the derivative with respect to $\tau$ is represented by a dot
and the derivative with respect to $\sigma$ by a prime. Also, one
has to impose the boundary conditions
\be
\dot{{\bf q}}=0,\;\;b=c=0,
\ee
on the solid lines. These boundary conditions can be implemented
 by introducing Lagrange multipliers ${\bf y}$, $\bar{b}$ and 
$\bar{c}$ and adding suitable terms to the action. Since ghosts will
not be needed in what follows, from now on, we will drop them. The action
with the boundary conditions included, but without the ghosts, reads
\be
S=\intst \left(-\frac{1}{2}{\bf q}'^{2}+\rho\, {\bf y}\cdot \dot{{\bf q}}
\right).
\ee
Here the field $\rho$ is a delta function on the boundaries and it
vanishes in the bulk: it is inserted to ensure that the boundary condition
is imposed only on the boundaries. However,  with this
insertion, the part of the integral over ${\bf y}$ that has support in
the bulk diverges, since the integrand over this region is independent
of ${\bf y}$. To avoid this problem, we add a Gaussian term to the
action which cuts off the divergence:
\bq
S&\rightarrow& S+ S_{g.f},\nonumber\\
S_{g.f}&=& \intst \left(-\frac{1}{2} \alpha^{2}\bar{\rho}\,
{\bf y}^{2}\right),
\eq
where $\alpha$ is a constant and $\bar{\rho}$ is complimentary to
$\rho$: it vanishes on the boundaries and it is equal to one
 everywhere else. It was pointed out in [5,6] that this  can be
thought of as a gauge fixing term. In its absence, the action is
invariant under the gauge transformation
$$
{\bf y}\rightarrow {\bf y}+ \bar{\rho}\, {\bf z},
$$
where ${\bf z}$ is an arbitrary function of the coordinates. It may seem
that we have introduced a new parameter $\alpha$ into the model, but
we will see in section 6 that this new parameter can be absorbed into
the definition  of the cutoff parameters that will be
needed shortly.

It is possible to give an explicit construction for the fields $\rho$ and
$\bar{\rho}$ in terms of a fermionic field on the world sheet. To see
how this works, it is best to discretize the $\sigma$ coordinate into
segments of length $a$. This discretization is pictured in Fig.3 as
a collection of parallel line segments, some solid and some dotted, spaced 
distance $a$ apart.
\begin{figure}[t]
\centerline{\epsfig{file=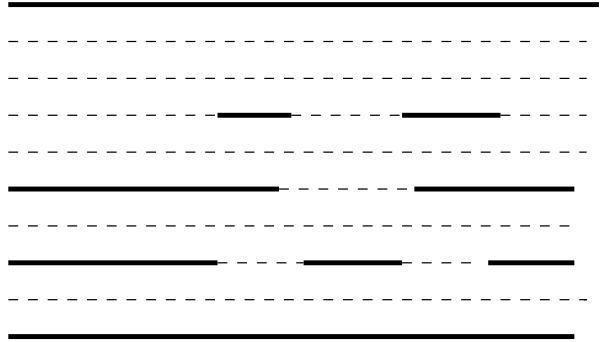, width=8cm}}
\caption{Solid and dotted lines}
\end{figure}
 The boundaries are marked by the solid lines, and
the bulk is filled with the dotted lines. Associated with these lines,
there are a two
component fermion field $\psi_{i}(\sigma_{n},\tau)$ and also its adjoint
$\bar{\psi}_{i}$, where, $\sigma_{n}=n a$, is the discretized $\sigma$
coordinate. The field $\bar{\psi}_{1}$ creates
 a dotted line and $\bar{\psi}_{2}$
a solid line out of vacuum, and $\psi_{1,2}$ annihilate these lines.
 $\rho$ and $\bar{\rho}$ can now be written as
\be
\rho=\frac{1}{2}\bar{\psi}(1-\sigma_{3})\psi,\;\bar{\rho}=\frac{1}{2}
\bar{\psi}(1+\sigma_{3})\psi,
\ee
and the fermionic action is given by
\bq
S_{f}&=&\sum_{n}\int d\tau\left(i\bar{\psi}\dot{\psi}-g\,\bar{\psi}
\sigma_{1}\psi\right)_{\sigma=n a}\nonumber\\
&\rightarrow&\intst\left(i\bar{\psi}\dot{\psi}-g\, \bar{\psi}
\sigma_{1}\psi\right).
\eq

Here, the first line represents the action in terms of the discretized
fermions and the second line the corresponding continuum limit. The first
term in the action corresponds to free fermion propagation along the
$\tau$ direction, and the second term to interaction taking place
 when there is 
transition from a dotted line to a solid line or vice versa. We will 
work with both the discrete and the continuum pictures; the discrete
version will be particularly useful in regulating the model. Thus the
 parameter $a$ will serve as one of our two cutoffs. It is important
to notice that there is a change in the normalization of the fermion
field in passing from the discrete to the continuum picture. This is
because there is a factor of $a$ involved in converting a sum into an
 integral:
$$
a\sum_{n}\rightarrow \int d\sigma,
$$
which means that the fermion fields should be scaled as
\be
\psi\rightarrow \frac{1}{\sqrt{a}}\psi,\;\;
\bar{\psi}\rightarrow \frac{1}{\sqrt{a}}\bar{\psi},
\ee
in going over to the continuum normalization.

As we mentioned earlier, to reproduce all Feynman graphs, one has to
sum (integrate) over all possible positions of the solid lines (boundaries).
It can readily been seen that summing over the two components of the
fermion field at each point on the world sheet is equivalent to summing
over all possible positions of the boundaries. Therefore, the introduction
of the fermionic field enables one, at least formally, to sum over all
planar Feynman graphs of the massless $\phi^{3}$ theory, and
 the following world sheet
action , gotten by adding the eqs.(4),(5) and(7),  provides a
compact expression for this sum:
\be
S=\intst \left(-\frac{1}{2}{\bf q}'^{2}+\rho\, {\bf y}\cdot\dot{{\bf q}}
-\frac{1}{2}\alpha^{2}\bar{\rho}\,{\bf y}^{2}+i\bar{\psi}\dot{\psi}
-g\,\bar{\psi}\sigma_{1}\psi\right),
\ee
where $\rho$ and $\bar{\rho}$ are given by eq.(6).

There is, however, a problem with the above action:
It fails to reproduce the prefactor $1/(2\,p^{+})$ in eq.(1) for
the propagator, unless, $g$, instead of being a constant, is allowed
to depend on $\rho$.
 In [6], it was shown how to take this dependence into account
within the meanfield approximation. We will ignore this problem, since
 we will see later that this complication does not arise in
summing over strings.

Finally, we would like to discuss the behaviour of various fields under
the scaling of coordinates, which is intimately connected with Lorentz
invariance. In the light cone setup we have, the only Lorentz transformation
that is still manifest is generated by the boost along the singled out
direction labeled by $1$. Under this boost, $x^{+}$ and $p^{+}$ scale as
$$
x^{+}\rightarrow u\, x^{+},\;\;p^{+}\rightarrow u\,p^{+},
$$
where $u$ parametrizes the boost. If under this scaling, the fields
transform as
\bq
{\bf q}(\sigma,\tau)\rightarrow& {\bf q}(u\sigma,u\tau),\;\;
{\bf y}(\sigma,\tau)\rightarrow& {\bf y}(u\sigma,u\tau),\nonumber\\
\psi(\sigma,\tau)\rightarrow& \sqrt{u}\,\psi(u\sigma,u\tau),\;\;
\bar{\psi}(\sigma,\tau)\rightarrow& \sqrt{u}\,\bar{\psi}(u\sigma,
u\tau),
\eq
then the action given by eq.(9) is invariant except for the interaction and
the gauge fixing terms. These two terms become invariant, at least
formally, only if we also require that
\be
g\rightarrow u\, g,\;\;\alpha^{2}\rightarrow u\, \alpha^{2}.
\ee
It is somewhat unusual to require constants in an action to transform,
and one may worry that Lorentz invariance is in danger. We will see
later on how this problem is resolved.

This finishes the review of the massless $\phi^{3}$ theory on the
world sheet. We will not review the calculation of the ground state of
this model in the mean field approximation given in reference [6], since
in any case, the mean field method will be developed in the context
summation of planar string graphs in the next sections.

\vskip 9pt

\noindent{\bf 3. String Summation}

\vskip 9pt

We start with the open bosonic string in the light cone picture,
 with  $U(N)$ Chan-Paton factors. Taking the large $N$ limit picks
the planar graphs. For simplicity,  the length of the world
sheet in the $\tau$ direction is taken to be infinite, and  periodic
boundary conditions at $\sigma=0$ and $\sigma=p^{+}$
 are imposed, where $p^{+}$
is the total $+$ component of the momentum flowing into the world sheet.
We will also use $p^{+}$ freely to denote the $+$ momentum flowing
into individual strings; it will be clear from the context which
$p^{+}$ is meant.
The above setup has the advantage of being traslation invariant along both
the $\sigma$ and $\tau$ directions, which simplifies the subsequent
calculations considerably.  Fig.2, which pictured a $\phi^{3}$
field theory graph, applies equally well a general planar
open string graph, with the boundaries of individual propagating
strings again marked by the solid lines (see, for example, [12]).
Therefore, the fermionic action introduced in the last section for
summing over the graphs of the $\phi^{3}$ theory, works just as well
for summing over planar open string graphs. Of course,
 there are some differences between field theory and string theory
pictures. For example,
 the action $S_{0}$ for the free string propagator is now given by
\be
S_{0}= \intst \left(\frac{1}{2}\dot{{\bf x}}^{2}- \frac{1}{2 \beta^{2}}
{\bf x}'^{2}\right).
\ee
Comparing with eq.(2), we see that  the transverse momentum ${\bf q}$
is replaced by the transverse position ${\bf x}$, and there an additional
term involving a derivative with respect to time.
 We have also introduced an adjustable slope $\beta/\pi$ since we are
ultimately interested in the zero slope, or the field theory limit.
 Furthermore, the boundary condition on the solid lines
is now different: The Dirichlet condition (eq.(3)) is  replaced by
the Neumann condition
\be
{\bf x}'=0.
\ee
The fermionic part of the action is unchanged. The total action is
therefore given by
\be
S=\intst \left(\frac{1}{2}\dot{{\bf x}}^{2}-\frac{1}{2 \beta^{2}}
{\bf x}'^{2}+\rho\,{\bf y}\cdot {\bf x}' -\frac{1}{2}\alpha^{2}
\bar{\rho}\,{\bf y}^{2}+i\bar{\psi}\dot{\psi}-g\,\bar{\psi}\sigma_{1}
\psi\right).
\ee
Here, just as in eq.(9), the Lagrange multiplier ${\bf y}$ enforces the
boundary conditions, and the term proportional to $\alpha^{2}$ is
inserted to cut off the divergent integral over ${\bf y}$ in the bulk.
This world sheet action then provides a compact expression for
the sum of planar open string graphs.
We note that, unlike in the case of $\phi^{3}$ field theory,
where a prefactor in the propagator was missing ,
the string propagator of eq.(12) is exact. Therefore, the complication
discussed following eq.(9) does not arise, and $g$ can simply be taken
to be a constant. An additional simplification is the absence of the
ghost fields $b$ and $c$ (see eq.(2)).
 From this point of view, the sum over string
graphs looks considerably simpler than the sum over field theory graphs.

In order to facilitate the comparison with field theory, we would like
to convert (14) into a form as close to eq.(9) as possible. This will involve
 a duality transformation which makes the following interchanges:
\be
{\bf x}\leftrightarrow {\bf q},\;\;{\bf x}'=0\leftrightarrow 
\dot{\bf q}=0.
\ee
The first step of the duality transformation is to integrate over
${\bf x}$ in eq.(14):
\bq
S&\rightarrow& S_{f}+ S_{g.f}+
 \frac{1}{2} i D\, Tr\ln\left(\beta^{2}\partial_{\tau}^{2}
-\partial_{\sigma}^{2}\right)\nonumber\\
&&\nonumber\\
&+&\frac{\beta^{2}}{2}\intdbl\,
 G(\sigma \tau,\sigma' \tau')\,\partial_{\sigma}(\rho {\bf y})_{\sigma\tau}
\,\partial_{\sigma'}(\rho {\bf y})_{\sigma'\tau'},\nonumber\\
&&
\eq
where $S_{f}$ and $S_{g.f}$ are given by eqs.(7) and (5),
 and the Green's function $G$ satisfies
\be
(\beta^{2} \partial_{\tau}^{2}- \partial_{\sigma}^{2})
G(\sigma\tau,\sigma'\tau')=\delta(\sigma-\sigma')\delta(\tau-\tau').
\ee
We note that, because of the translation invariance on the world sheet,
the Green's function depends only on the coordinate differences
$\sigma-\sigma'$ and $\tau-\tau'$.

Next, we integrate the $\sigma$ derivatives by parts, and use the
translation invariance of the Green's function and the defining equation
(17) to arrive at
\bq
S&=&S_{f}+S_{g.f}+
 \frac{1}{2} i D\, Tr\ln\left(\beta^{2}\partial_{\tau}^{2}
-\partial_{\sigma}^{2}\right)+ \frac{\beta^{2}}{2}\intst \rho^{2}\,
{\bf y}^{2}\nonumber\\
&&\nonumber\\
&+&\frac{\beta^{4}}{2}\intdbl\, G(\sigma\tau,\sigma'\tau')
\,\partial_{\tau}(\rho {\bf y})_{\sigma\tau} \,\partial_{\tau'}
(\rho {\bf y})_{\sigma'\tau'}.
\eq
 In order to have a sensible $\beta\rightarrow 0$
limit in this equation, we first scale ${\bf y}$ by
$$
{\bf y}\rightarrow {\bf y}/\beta^{2},
$$
and rewrite it by introducing an auxiliary variable ${\bf q}$:
\bq
S&=&S_{f}+S_{g.f}+ \frac{1}{2\beta^{2}}\intst \rho^{2}\,
{\bf y}^{2}\nonumber\\
&+&\intst \left(\frac{1}{2}\beta^{2}\dot{\bf q}^{2}
-\frac{1}{2} {\bf q}'^{2}+ \rho\,{\bf y}\cdot \dot{\bf q}\right).
\eq

This action is quite similar to the world sheet action for $\phi^{3}$
(eq.(9)) and in fact coincides with it in the zero slope
 limit $\beta\rightarrow 0$, except for the term
$$
S'=\frac{1}{2 \beta^{2}}\intst\, \rho^{2}\,{\bf y}^{2},
$$
which blows up in this limit. We will now argue that this term should be
absent. In fact, our starting point, eq.(14), was not quite correct; in
the action $S$, the term $S'$ should have been dropped.
This point is perhaps made clearer by considering an electrostatic analogy.
The Lagrange multiplier ${\bf y}$ which enforces ${\bf x}'=0$
 on the boundaries can be thought of as a line charge induced by
 the boundary condition. It is easy to see that $S'$ is 
 the (divergent) electrostatic self energy of the line charge in question.
On the other hand,
the Lagrange multiplier ${\bf y}$ was introduced solely to enforce the Neumann
boundary conditions; the induced electrostatic energy is an additional
boundary term which is absent in the usual treatment
of the open string. It should therefore be dropped.
 As a further check,  consider the configuration
 with two eternal boundaries at $\sigma=0$ and $\sigma=p^{+}$:
For this simple case, $\rho=\delta(\sigma)+\delta(\sigma-p^{+})$ and,
after scaling ${\bf y}$ by $1/\beta^{2}$, the last term in eq.(18)
reduces to
\be
S\rightarrow \frac{1}{2}\intdbl\, G(\sigma\tau,\sigma'\tau')
\,\partial_{\tau}({\bf y})_{\sigma\tau} \,\partial_{\tau'}
({\bf y})_{\sigma'\tau'}.
\ee
The functional integral over ${\bf y}$ can be done,
 and after a simple calculation
which we will not present here, the correct free propagator for the open
 string
is reproduced. This justifies the dropping of $S'$ from the world sheet
 action.  

\vskip 9pt

\noindent{\bf 4. The Meanfield Calculation}

\vskip 9pt

The meanfield calculation we are going to present here is very similar to the
treatment for the $\phi^{3}$ theory given in [2-6]. There are, however, some
important differences, which we will point out as we go along. 
 The starting point is the world sheet
action, derived at the end of the last section, which we write in full:
\be
S=\intst \left(\frac{1}{2}\beta^{2}\dot{{\bf q}}^{2}-\frac{1}{2}{\bf q}'^{2}
+\rho\,{\bf y}\cdot\dot{{\bf q}} -\frac{1}{2} \alpha^{2} \bar{\rho} {\bf y}^{2}
+i\bar{\psi}\dot{\psi}- g\,\bar{\psi}\sigma_{1}\psi\right).
\ee
Ultimately we will be interested in the $\beta\rightarrow 0$
(zero slope) limit,
which will get us back to field theory, but we will study this limit
only within the framework of the meanfield method. The reason for this
indirect approach is that if one naively sets $\beta=0$ in the above 
expression for $S$, one does not quite get the correct field theory
result. For example, the ghost fields $b$ and $c$ are missing and
also there is the problem of the missing prefactor in the propagator
discussed following eq.(9). Later on, we will argue that the naive
$\beta\rightarrow 0$ limit can be quite singular, but that
  a smooth limit can be defined within the
framework of the mean field method.

It is convenient to view the mean field approximation as the large $D$
limit of the field theory defined on the world sheet, where $D$ is the
number of transverse dimensions. We hasten to add that this is merely
a convenient way of doing the correct bookkeeping; one can set $D$ to
any desired value at the end of the calculation. The idea is to cast
the action into a form proportional to $D$ and take the large $D$ limit
by the saddle point method. This is the standard way of solving for the
anologous large $N$ limit of so called vector models [13]. Following [5,6],
we introduce the extra term $\Delta S$ in the action:
\bq
S&\rightarrow& S+ \Delta S,\nonumber\\
\Delta S&=&\intst \left(\kappa_{1}(D\, \lambda_{1}- {\bf y}\cdot
\dot{{\bf q}})+\frac{1}{2}\kappa_{2}(D\,\lambda_{2}-{\bf y}^{2})\right).
\eq
Integrating over $\kappa_{1,2}$, all we have done is to rename the composite
fields ${\bf y}\cdot \dot{{\bf q}}$ and ${\bf y}^{2}$ as $D\lambda_{1}$
and $D\lambda_{2}$. The factors of $D$ are natural since each composite field
is the sum of $D$ terms. The Gaussian integral over ${\bf y}$ is easily
done, with the result,
\bq
S&+&\Delta S\rightarrow S_{1}+S_{2}+S_{3},\nonumber\\
S_{1}&=& \intst \left(\frac{1}{2}(\beta^{2}+\kappa_{1}^{2}/\kappa_{2})\,
\dot{{\bf q}}^{2} -\frac{1}{2}{\bf q}'^{2}\right),\nonumber\\
S_{2}&=&D \intst \left(\kappa_{1}\lambda_{1}+\frac{1}{2}\kappa_{2}
\lambda_{2}\right),\nonumber\\
S_{3}&=&\intst \left(i\bar{\psi}\dot{\psi}- g\,\bar{\psi}\sigma_{1}
\psi+ \frac{D}{2}(\lambda_{-}\bar{\psi}\psi-\lambda_{+}\bar{\psi}
\sigma_{3}\psi)\right),
\eq
where, we have defined,
$$
\lambda_{\pm}=\pm \lambda_{1}+\frac{1}{2}\alpha^{2}\lambda_{2}.
$$ 

Some of the terms in this equation can be further simplified. We observe that
the operator $\bar{\psi}\psi$ represents the local fermion density.
Since there is always one fermion on each horizontal line, independent
of whether it is dotted or solid, one can set this operator equal to unity
 in the picture where the $\sigma$ coordinate is discretized.
 On the other hand, in
 the continuum normalization, taking into account
the scaling given by eq.(8), one can instead set
\be
\bar{\psi}\psi=1/a.
\ee
After this substitution, $\lambda_{-}$ becomes a Lagrange multiplier,
enforcing the constraint
\be
\kappa_{2}=\alpha^{2}(\kappa_{1}+1/a).
\ee
With these simplifications (eqs.(24) and (25)),
 the world sheet action becomes,
\bq
S&=&\intst \Big(\frac{1}{2} A^{2} \dot{{\bf q}}^{2}-\frac{1}{2}
{\bf q}'^{2}+\lambda(\kappa+ 1/(2 a))\nonumber\\
&+&i\bar{\psi}
\dot{\psi}- D g\,\bar{\psi}\sigma_{1}\psi- \frac{D}{2}\lambda\,
\bar{\psi}\sigma_{3}\psi\Big),
\eq
where,
\be
A^{2}=\beta^{2}+\frac{\kappa^{2}}{\alpha^{2}(\kappa+1/a)},
\ee
and we have scaled the coupling constant by $D$,
\be
g\rightarrow D\,g,
\ee
in order to have an action that is proportional to $D$
in the large $D$ limit. Also, to simplify writing,
we have set
$$
\kappa_{1}=\kappa,\;\;\lambda_{+}=\lambda.
$$
It is important to note that after summing over strings, the slope
parameter changed from $\beta$ for the free string to the dynamical
variable $A$. We will later compute the ground state
expextation value of $A$, and show that it can differ from $\beta$.

Before closing this section, we would like to stress that the parameters
so far introduced that define the model
 are in general cutoff dependent bare parameters. We already
have one cutoff $a$, the spacing of the grid along the $\sigma$ direction,
and  $a'$, another grid spacing along the $\tau=x^{+}$ direction
will soon be needed in order to regulate the integral over ${\bf q}$.
How are the renormalized parameters, which should stay finite as the
cutoffs are removed by letting
$$
a\rightarrow 0,\;\;a'\rightarrow 0,
$$
to be defined? We will not address the problem of renormalization
in any detail here\footnote{ See [14] for an investigation of
renormalization and Lorentz invariance in the light cone formulation.},
but one obvious condition is to demand that the renormalized parameters
be invariant under the scaling transformation discussed at the end of 
the section 2. Since the scale transformation is the same as a
special Lorentz transformation, this is clearly  necessary
for Lorentz invariance. The idea is then to define new
 scale invariant parameters by multiplying them with appropriate powers
of $a$ and $a'$. The slope $\beta$, which has the dimension of inverse
mass squared, is already scale invariant. We also note
that  $a$ and $a'$
have the scaling properties (eq.(10)) and the dimensions of $p^{+}$ and
$x^{+}$ respectively, so that the ratio
\be
a/a'=m^{2},
\ee
is scale invariant and has the dimension of mass squared. We will
 hold this ratio finite and fixed in the limit of $a$ and $a'$
going to zero. Therefore, there is only one cutoff, say $a$, and a
mass parameter $m$ in the problem. Eventually, we will consider the
limit $\beta\rightarrow 0$ limit, and $m$ will then be the only mass
left in the model to set the mass scale.

In addition to $\beta$, $a$ and $a'$, there are two more constants
in the problem: The coupling constant $g$ and the gauge fixing parameter
$\alpha$. We trade them for scale invariant constants $\bar{g}$ and
$\bar{\alpha}$ by defining 
\be
\bar{\alpha}^{2}=\frac{\alpha^{2} a'^{2}}{a},\;\;
\bar{g}=\frac{g a'}{\pi},
\ee
where the factor of $\pi$ is introduced for later convenience.
There was actually an ambiguity in the definition of the barred
constants because of the availability of the scale invariant parameter
$m$; we fixed this ambiguity by requiring $\bar{g}$ and $\bar{\alpha}$
to be dimensionless.
We shall see in the next section that the slope of the interacting
string is expressible in terms of these new constants, without any
explicit dependence on the cutoff. This will provide some justification
for calling them renormalized constants.

\vskip 9pt

\noindent{\bf 5. The Ground State Of The Model}

\vskip 9pt

In this section,  the ground state of the model will be determined 
 by minimizing the energy of the system.
So far, everything has been exact: No approximations were made, for
example, in deriving eq.(23). Of course, we are unable to do an
exact calculation, so
 to make progress, we have to appeal to 
 the large $D$ limit. In this limit, the fields $\kappa$
and $\lambda$ are  treated as classical quantities, to be calculated
by the saddle point method. On the other hand, ${\bf q}$, $\psi$ and
$\bar{\psi}$ are still full quantum fields, to be
integrated over functionally. In other words, in the leading large
$D$ limit, $\kappa$ and $\lambda$ are to be replaced by their
ground state expectation values
$$
\kappa_{0}=\langle \kappa \rangle,\;\; \lambda_{0}=\langle \lambda
\rangle.
$$
In order to avoid excessive notation, from now on,
 we will drop the subscript zero, so that $\kappa$ and $\lambda$
will stand for the expectation values of these fields.

At this point, translation invariance along both the $\sigma$ and
$\tau$ directions comes in handy. It allows us set $\kappa$ and
$\lambda$ equal to constants independent of the coordinates $\sigma$
and $\tau$. This means that $A$ is also independent of the coordinates,
and therefore the integrals over ${\bf q}$, $\psi$ and $\bar{\psi}$ in the
 action $S$ of eq.(26) can be done explicitly.
 Instead of evaluating a given action $S$ directly,
 we find it more convenient
to compute the corresponding energy $E$  and 
express $S$ in terms of $E$ by means of
$$
S= -i\tfi E,
$$
where $\tfi$ is the (infinite) time interval.
For example, the result of carrying out the integral over ${\bf q}$
in eq.(26) can be expressed as
\be
S_{1}= \frac{i D}{2}\,Tr\ln\left(A^{2}\partial_{\tau}^{2}
-\partial_{\sigma}^{2}\right)= -i\tfi E^{(0)},
\ee
where $E^{(0)}$ is  the zero point energy of the free
string. Similarly, the result of doing the $\psi$ and $\bar{\psi}$
integrations can be expressed in terms of the fermionic energy
$E_{f}$. Adding up, the total energy corresponding to $S$ in
eq.(26) is given by
\be
E= E^{(0)}+ E_{f}- p^{+}\lambda\, (\kappa+1/(2 a)).
\ee
One point should be clarified here. What we have called the energy 
$E$ is really the light cone energy $p^{-}$. Because of the periodic
boundary conditions, the total transverse momentum is zero, and  
the invariant mass squared $M^{2}$ of the system is given by
$$
M^{2}= p^{+}\,E.
$$

The next step is to evaluate $E^{(0)}$ and $E_{f}$. Since
 $E^{(0)}$ is well known from the
standard calculation of the Casimir effect, we only remind the reader
of the steps involved. The regulated zero point energy is given by
\be
E^{(0)}=D \sum_{0}^{\infty}E_{k}\,\exp(- E_{k}/\Lambda),
\ee
where $E_{k}$ is the zero point energy of the $k$'th SHO mode,
\be
E_{k}=\frac{2\pi k}{p^{+}},
\ee
and we have introduced an exponential regulator with the 
parameter $\Lambda$. The leading two terms in the limit
$\Lambda\rightarrow \infty$ are given by
\be
E^{(0)}\rightarrow D\left( \frac{A\,\Lambda^{2}\,p^{+}}{2\pi}- \frac{\pi}
{3 A\, p^{+}}\right).
\ee
Of course, any other smooth regulator that depends only on
$E_{n}$ gives the same result.

The regulator $\Lambda$ acts as a cutoff in energy; it is
related to $a'$, the spacing of the grid
 in the conjugate variable $\tau$, by
$$
a'=\frac{2\pi}{\Lambda},
$$
and replacing $\Lambda$ by $a'$ in eq.(35) gives
\be
E^{(0)}=2 \pi D\, p^{+} A/(a')^{2},
\ee
where we have kept the cutoff dependent term and dropped the finite one.
In calculating the Casimir effect, one does the opposite: The cutoff
dependent term is subtracted and the finite term is kept. Here, this term,
through A, depends on the dynamical variable $\kappa$ (eq.(27)), and there is
no way to cancel it by introducing a constant counter term, independent
of the dynamical variables. In fact, since the cutoff dependent term dominates
over the finite term, we have dropped the latter.
  
 The fermionic energy $E_{f}$ is evaluated by diagonalizing
the corresponding Hamiltonian
\bq
H_{f}&=&\sum_{n}H^{(n)},\nonumber\\
H^{(n)}&=& D \left(\frac{1}{2}\lambda\, \bar{\psi}\sigma_{3}\psi
+ g\, \bar{\psi}\sigma_{1}\psi\right)_{\sigma=n a},
\eq
which has been regulated by discretizing $\sigma$. We have already remarked
that $\lambda$ (and of course, also $g$) is a constant, and therefore,
$H^{(n)}$ reduces to a constant two by two matrix in the two
dimensional space spanned by $\bar{\psi}_{i}$:
\be
H^{(n)}\rightarrow \left(\begin{array}{cc} \lambda/2 & g \\
g & -\lambda/2 \end{array}\right).
\ee 
The two eigenvalues of this matrix ,
$$
E_{\pm}=\pm \frac{1}{2}\sqrt{\lambda^{2}+ 4 g^{2}},
$$
have to be multiplied by the number of points forming the $\sigma$ grid,
$p^{+}/a$, in order to obtain the total fermionic energy:
\be
E_{f,\pm}= \frac{p^{+}}{2 a}\sqrt{\lambda^{2}+4 g^{2}}.
\ee
Notice that the fermionic energy has two possible values. Clearly,
the choice $(-)$ corresponds to the ground state, but we will also be
interested in the other possibility.

Putting together eqs.(32),(35) and (39), we have the following expression
for the total energy:
\bq
E_{\pm}&=& D p^{+}\Bigg(\frac{2\pi}{a'^{2}}\sqrt{\beta^{2}+\frac{\kappa^{2}}
{\alpha^{2}(\kappa+1/a)}}\pm \frac{1}{2 a}\sqrt{\lambda^{2}+ 4 g^{2}}
\nonumber\\
&-&\frac{1}{2}\lambda(2\kappa+1/a)\Bigg).
\eq
As  explained earlier, we are looking for the saddle point
of this expression in the variables $\lambda$ and $\kappa$.
The saddle point satisfies,
$$
\frac{\partial E_{\pm}}{\partial\lambda}=0,\;\;
\frac{\partial E_{\pm}}{\partial \kappa}=0.
$$
The first equation determines $\lambda$:
$$
\lambda=\pm g\frac{1+2 a \kappa}{\sqrt{-a\kappa- a^{2}\kappa^{2}}},
$$
and using this result,  the $\lambda$ dependence of
the energy can be eliminated,
 leaving it as a function of only $\kappa$. Before
writing down the result,
it is convenient to replace $\kappa$ by a cutoff independent and
dimensionless new variable $x$ through
\be
x=- a\kappa,
\ee
and the constants $\alpha$ and $g$ by their cutoff
independent counterparts $\bar{\alpha}$ and $\bar{g}$ through eq.(30).
After these substitutions, the expression for the energy is given
by
\be
E_{\pm}=\frac{2\pi D\,p^{+}}{a a'}\left(\sqrt{\beta^{2} m^{4}
+x^{2}/(1-x)}\pm \bar{g}\sqrt{x- x^{2}}\right).
\ee

We pause briefly to discuss the physical significance of $x$. By
computing the eigenvectors of the matrix (38), it is easy to
show  that [6],
\be
\frac{1}{2}\langle \bar{\psi}(1-\sigma_{3})\psi\rangle
=\langle \rho\rangle=x,\;\;\frac{1}{2}\langle\bar{\psi}
(1+\sigma_{3})\psi=\langle \bar{\rho}\rangle=1- x,
\ee
where $\langle \rangle$ represents the ground state expectation
value. This is in the discretized version of the world sheet; in the 
continuum version, $x$ and $1-x$ should be replaced by $x/a$
and $(1-x)/a$. Therefore, in the discrete version, $x$ is the average
probability of  finding a spin down fermion on the world sheet. By
the definition, this is the same as the average
probability of finding a solid line. Conversely, $1-x$ is the
average probability of finding a dotted line. From this probability
interpretation, it is clear that
\be
0\leq x\leq 1.
\ee  
We should emphasize that, for the probability to be well defined,
it was important to have a discretized world sheet, with the grid
spacing $a$ kept fixed.
 
The next step is to find the minima of $E_{\pm}$ as a function of
$x$. This is easy in the case of $E_{+}$: It has a minimum at
$x=0$, with the value
\be
E_{+}= \frac{2\pi D\,p^{+} \beta m^{2}}{a a'}.
\ee
The true minimum is, of course, the minimum of $E_{-}$. The value of $x_{m}$
that minimizes $E_{-}$ cannot be found analytically, but one can get
approximate answers in the two interesting limits: $\bar{g}\ll 1$ (weak
coupling), and $\bar{g}\gg 1$ (strong coupling). We have also to distinguish
between two cases, depending on whether the initial slope is non-zero
($\beta m^{2}\neq 0$), or it is zero
($\beta m^{2}=0$). Taking $\beta m^{2}\neq 0$ and
$\bar{g}\ll 1$, to leading order in $\bar{g}$, the minimum is given by,
\bq
x_{m}&\approx& \left(\frac{\beta m^{2}\bar{g}}{2}\right)^{2/3},
\nonumber\\
 E_{-,m}&\approx&\frac{2\pi D\,p^{+}}{a a'}\left(
 \beta m^{2}- (2)^{-1/3}(\bar{g})^{4/3}
(\beta m^{2})^{1/3}\right).
\eq
We see that, in the weak coupling limit, $x_{m}$ is small and the minimum
of $E_{-}$ is less then the minimum of $E_{+}$, as expected. On the other
hand, in the strong coupling limit, $\bar{g}\gg 1$, $x_{m}$
approaches $1/2$:
\bq
x_{m}&\approx& 1/2- \frac{3}{4 \bar{g}}\left(\beta^{2} m^{4}+1/2\right)
^{-1/2},\nonumber\\
E_{-,m}&\approx& -\frac{\pi D\,p^{+}}{a a'}\bar{g}.
\eq

Now let us consider the case of zero slope for the free string,
$\beta m^{2}=0$. in the weak coupling limit, the minimum is given by
\be
x_{m}\approx \bar{g}^{2}/4,\;\;E_{-,m}\approx - \frac{\pi D\,p^{+}}
{2 a a'} \bar{g}^{2},
\ee
and in the strong coupling limit by
\be
x_{m}\approx 1/2- \frac{3}{2\sqrt{2}\bar{g}},\;\;
E_{-,m}\approx - \frac{\pi D\,p^{+}}{2 a a'} \bar{g}^{2}.
\ee

From the above results, it is clear that the cases of non-zero and
zero initial  slope are qualitatively similar. In both cases,
$x_{m}$ ranges from zero to $1/2$ as
the coupling constant $\bar{g}$ varies from zero to infinity. The
function,  
$$
f(x)=\frac{a a'}{2\pi D\,p^{+}}\,E_{-}=\sqrt{\beta^{2} m^{4}
+x^{2}/(1-x)}-\bar{g}\sqrt{x-x^{2}},
$$
is plotted against $x$
for $\beta m^{2}=10$, $\bar{g}=1$ and for $\beta m^{2}=0$,
$\bar{g}=1$, in Figs.4 and 5  respectively.
\begin{figure}[t]
\centerline{\epsfig{file=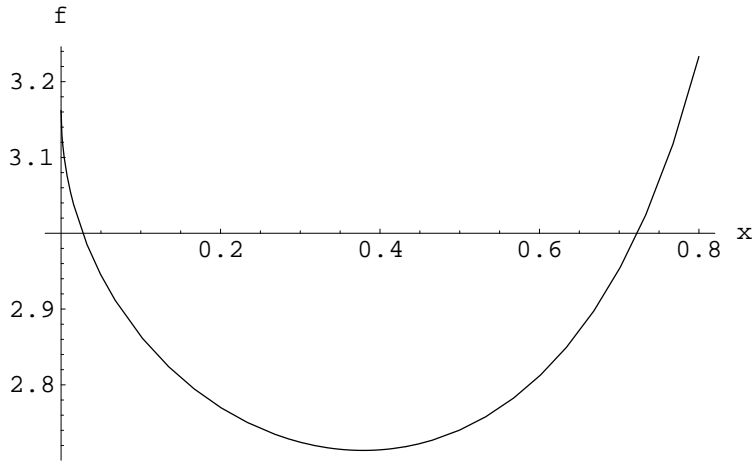, width=10cm}}
\caption{The plot of $f(x)$ against $x$ for $\beta m^{2}=10$
and $\bar{g}=1$}
\end{figure}
\begin{figure}[t]
\centerline{\epsfig{file=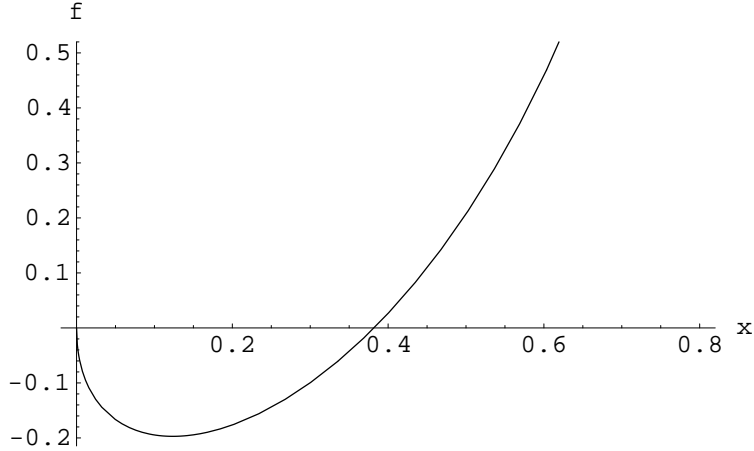, width=10cm}}
\caption{The plot of $f(x)$ against $x$ for $\beta m^{2}=0$ and
$\bar{g}=1$}
\end{figure} 

\vskip 9pt

\noindent{\bf 6. Discussion Of The Results}

\vskip 9pt

In the last section, we have seen that:\\
1) There are two saddle points of the model, with ground state energies
$E_{+}$ and $E_{-}$. The true ground state of the model corresponds to
 $E_{-}$, which is always less than $E_{+}$. 
 We will call the
first solution the $(+)$ phase and the second one the $(-)$ phase.\\
2) The minimum of $E_{+}$ is realized at $x_{m}=0$, whereas the minimum
of $E_{-}$ is at some value of $x_{m}$ that satisfies
$$
0<x_{m}<1/2.
$$
3) These statements are true for both finite initial slope $\beta$, and also
in the limit $\beta\rightarrow 0$.

We argued in the last section, following eq.(43), that $x$ represents
the average probability of finding a solid 
line on the discretized world sheet. Since
the solid lines represent the boundaries, in effect, $x$ measures the average
density of the string boundaries. For
 the perturbative sum over strings that we started with, we expect
$x_{m}$ to be zero, since at each order of perturbation, the boundaries
form a set of measure zero. It is then natural 
to identify the solution corresponding to the $(+)$ phase, whose minimum is at
$x_{m}=0$, with the perturbative string sum. In contrast, in the $(-)$ phase,
 where $x_{m}\neq 0$, the boundaries form a
dense set on the world sheet. We will call this phenomenon, which cannot
be realized in any finite order of perturbation, the condensation of
string boundaries. The order parameter that distinguishes between these
two phases is $x$, the expectation value of the composite operator
$\rho$ (eq.(43)). The statements made above are quite general, independent
of any approximation scheme. Without a dynamical calculation,
 however, we do not know which of the phases has the lower energy.
 We have seen above that, at least in the
mean field approximation, the $(-)$ phase,
 in which the boundaries have condensed, has the lower ground state energy.

At the end of the last section, we  studied the dependence of $x_{m}$
in the $(-)$ phase
on the coupling constant $\bar{g}$, and found that $x_{m}\neq 0$ for
all non-zero $\bar{g}$. It is at first surprising that condensation
of boundaries takes place even at small $\bar{g}$, but we have to
remember that the original coupling constant $g$ has already been
scaled by a factor of $D$ (eq.(28)). The mean field approximation is
based on the limit $D\rightarrow \infty$, and in this limit, even
small values of $\bar{g}$ may give rise to large values of $g$. On
the other hand, it is easy to understand what happens for large $\bar{g}$.
In this limit, the fermionic energy $E_{f}$ (second term in eq.(42)),
which is proportional to  $\bar{g}$,
dominates, and the minimum of
 this term is at $x_{m}=1/2$. The limiting value of $x_{m}=1/2$ is also
easy to understand: The large coupling constant limit energetically
favors a maximum density of string vertices. Since vertices convert a solid
line into a dotted line and vice versa, it is advantageous to flip
 between solid and dotted lines as often as possible. It is easy
to see that this corresponds to an equal density for the solid and dotted
lines, namely, $x=1/2$.

 It was pointed out earlier that, after the
summation over free strings is carried out,
the free string slope $\beta/\pi$ is replaced by the $A/\pi$ .  This is
a fluctuationg dynamical variable, but  we could define an average slope in
terms of the ground state expectation value of $A$. Replacing $\alpha$
and $\kappa$ in eq.(27) by $\bar{\alpha}$ and $x$ through (30) and (41)
 gives  
\be
\langle A^{2}\rangle=\frac{1}{m^{4}}\left(\beta^{2} m^{4}+\frac{x^{2}_{m}}
{\bar{\alpha}^{2}(1-x_{m})}\right).
\ee
For the $(+)$ solution with $x_{m}=0$, the average slope after the summation
is the same as the free string slope, which is consistent with the
 perturbative picture. On the other hand, the $(-)$ solution,
 with  $x_{m}\neq 0$, gives
   a slope larger than the free slope. So what emerges
 is a new string with a larger slope, and also, of course, with
a more complicated structure. We can also see from the above equation that
$\bar{\alpha}$ is a redundant parameter; one can absorb it into the
definition of $m$ by redefining one of the cutoff parameters $a$ or $a'$.

To  show that $A$ becomes a genuine dynamical variable, one has
to go to next to leading order in the large $D$ limit. We will not go into
the details here, since the calculation is almost identical to the analogous
one in the case of field theory, given in references [5,6]. The end result
is that, a kinetic energy term in the action,
\be
S^{(2)}=\frac{D\,\ln(2\pi p^{+}/a')}{8\pi\, (\langle A^{2}\rangle)^{3/2}}
\intst \left( \langle A^{2}\rangle\, (\partial_{\tau}A)^{2} -
(\partial_{\sigma}A)^{2}\right),
\ee 
for $A$ is generated, and so this variable becomes truly dynamical.
What happens here is quite similar to what happens in some other
two dimensional models: An auxiliary classical variable acquires 
kinetic energy term and becomes truly dynamical when one loop contributions
are taken into account [16,17].

Let us now consider the limit $\beta\rightarrow 0$.
 As noted earlier, this limit is rather
delicate in the $(+)$ phase; for example,
naively letting $\beta\rightarrow 0$  directly in 
 eq.(21) is not correct. It is easy to spot the problem:
In this phase, as $\beta$ and $x_{m}$ go to zero, so does $A$,
 and at $A=0$, the mean field method
is no longer applicable. What happens is that the non-leading
terms tend to become singular. For example,
the second term in eq.(35) for the zero point energy, which was neglected
compared to the first term, blows up at $A=0$. Also, non-leading order terms
in the large $D$ expansion of $S_{1}$ (eq.(31)) become singular in
the same limit, invalidating the expansion.
 This can perhaps be guessed  by setting $A=0$ directly
in $S_{1}$; the resulting expression loses its $\tau$ dependendence
and becomes ill defined.
 In contrast,  in the $(-)$ phase, 
$A$ stays finite as $\beta\rightarrow 0$, and the non-leading terms in the
meanfield (large D) expansion remain well-defined.

Summarizing the foregoing discussion, we  conclude that, in the limit of
the initial slope tending to zero:\\
1) The induced slope $A$ also goes to zero in the $(+)$ phase, which
causes the breakdown of the mean field method.\\
2) In contrast, $A$ remains non-zero in the $(-)$ phase, and there are 
no obvious problems with the mean field method.
Since $(-)$ is the energetically favored phase, we believe that this
is actually what happens.\\
3) The string slope becomes a fluctuating dynamical variable (eq.(51)).

The zero slope limit is commonly thought of as the field theory limit.
As $\beta\rightarrow 0$, the massive particles decouple, and one is
left with a field theory built out of the massless particles. If we
accept this picture, it follows that the summation of the field 
theory graphs has led to string formation. The sequence of the steps
in the reasoning is the following: Start with the sum of  planar open 
strings, and then take the zero slope limit. Order by order in the
perturbation expansion, we expect the string graphs to tend to the field
theory graphs. However, this is a limit rather difficult to define
cleanly in mathematical terms because of the existence of the 
tachyon, whose mass squared goes to $-\infty$ in this limit, if the
vector meson mass is fixed at zero.
 If, however, instead of first taking the zero slope limit,
  one reverses the order of the steps by
first summing over strings and then taking the zero slope limit, the
mean field calculation goes through smoothly, and the final induced 
slope is non-zero, signaling the formation of a new string.

\vskip 9pt

\noindent{\bf 7. Conclusions And Future Directions}

\vskip 9pt

In this article, we have applied the mean field method to the
sum of planar open bosonic string diagrams on the world sheet. After
a duality transformation, the problem was cast in a form very
similar to the problem of summing planar $\phi^{3}$ graphs [2-6],
and the techniques developed earlier could be applied here. The results
were also similar: The ground state of the system turned out to be
a condensate of the open string boundaries. As a result, a new string was
formed, with a slope greater than the initial slope. Even in the limit
of vanishing initial slope, the final slope remained non-zero.

We end by listing some remaining open problems. We would like to
identify the zero slope limit of the initial string with the field
theory of the zero mass vector particle, but the existence of the tachyon
makes this identification problematic. A future project is to apply
the methods developed here to the tachyon free superstrings.
 Another problem is the cutoff dependence of the
ground state energy given by eq.(42). In reference [6], a similar cutoff
dependence was cancelled by introducing a bare mass term for the $\phi$
field. Since here our starting point is already a string theory, we do
not have this freedom initially. We could, nevertheless, introduce
a counter term at the end. This then brings up the question of the 
finite part of the ground state energy left over after the cancellation
of the cutoff dependence. We remind the reader that in the usual treatment
of the bosonic string, this finite part is related to the intercept and 
it is not arbitrary. In the lightcone formulation, it is determined
by requiring Lorentz invariance [15]. This brings up another important
open problem; namely, the Lorentz invariance of the string that emerges
after the condensation of the boundaries. If the
intercept of this new string is determined by Lorentz invariance,
this would shed light on the question of tachyon condensation in the
open string [10]. We hope to address at least some of these problems
in the future.

\vskip 9pt

\noindent{\bf Note Added:}

\vskip 9pt

After finishing this paper, reference [18] was brought to my attention.
There is considerable overlap between this reference and the present
article. Both pieces of work tackle the problem of summing planar strings
on the world sheet using the mean field approximation, and both find
condensation of boundaries and a new string with increased slope. There
are, however, several important differences in the treatment of the
problem. For example, in this article, unlike in [18], the string boundary
conditions are imposed by means of Lagrange multipliers, and before applying
the mean field approximation, Neumann conditions are replaced by Dirichlet
conditions by means of a duality transformation. Also, some of our
conclusions differ: Reference [12] finds  a free string at the end
of the summation, whereas we find a more complicated string with a
fluctuating dynamical slope. Furthermore, in contrast to [12], starting
with zero slope (field theory), we find string formation. These
differences in the final result are probably due to the differences
in the treatment of the problem mentioned above.

\vskip 9pt

\noindent{\bf Acknowledgements}

\vskip 9pt

I would like to thank Peter Orland for bringing reference [18] to my
attention.
This work is supported by the Director, Office of Science, Office of
High Energy and Nuclear Physics, of the US Department of Energy under
Contract DE-AC02-05CH11231.

\vskip 9pt

\noindent{\bf References}

\vskip 9pt

\begin{enumerate}
\item K.Bardakci and C.B.Thorn, Nucl.Phys. {\bf B 626} (2002) 287,
hep-th/0110301.
\item K.Bardakci and C.B.Thorn, Nucl.Phys. {\bf B 652} (2003) 196,
hep-th/0206205.
\item K.Bardakci and C.B.Thorn, Nucl.Phys. {\bf B 661} (2003) 235,
hep-th/0212254.
\item K.Bardakci, Nucl.Phys. {\bf B 667} (2004) 354, hep-th/0308197.
\item K.Bardakci, Nucl.Phys. {\bf B 698} (2004) 202, hep-th/0404076.
\item K.Bardakci, Nucl.Phys. {\bf B 715} (2005) 141, hep-th/0501107.
\item C.B.Thorn, Nucl.Phys. {\bf B 637} (2002) 272, hep-th/0203167.
\item S.Gudmundsson, C.B.Thorn, T.A.Tran, Nucl.Phys.{\bf B 649} (2003)
3, hep-th/0209102.
\item C.B.Thorn, T.A.Tran, Nucl.Phys. {\bf B 677} (2004) 289,
hep-th/0307203.
\item For a review, see W.Taylor and B.Zwiebach, D-Branes, Tachyons
and String Field Theory, hep-th/0311017.
\item G.'t Hooft, Nucl.Phys. {\bf B 72} (1974) 461.  
\item M.B.Green, J.H.Schwarz, E.Witten, ``Superstring Theory, Vol.2'',
Cambridge University Press (1987).
\item For a review of the large N method, see M.Moshe, J.Zinn-Justin,
Phys.Rep. {\bf 385} (2003) 69, hep-th/0306133.
\item C.B.Thorn, Nucl.Phys.{\bf B 699} (2004) 427, hep-th/0405018,
D.Chakrabarti, J.Qiu, C.B.Thorn, hep-th/0602026.
\item P.Goddard, J.Goldstone, C.Rebbi, C.B.Thorn, Nucl.Phys. {\bf B 56}
(1973) 109.
\item D.Gross, A.Neveu, Phys.Rev. {\bf D 10} (1974) 3235.
\item A.D'Adda, M.Luscher, P.Di Vecchia, Nucl.Phys. {bf B 146} (1978) 63.
\item P.Orland, Nucl.Phys. {\bf B 278} (1986) 790.
\end{enumerate}

\end{document}